\documentclass[12pt]{article}
\usepackage{epsfig}
\usepackage{eucal}

\newfam\mbffam \def\mbf#1{{\bf#1}}

\font\tenscr=rsfs10

\font\twelvescr=rsfs10 scaled \magstep 1
\skewchar\tenscr='177
\newfam\scrfam \textfont\scrfam=\twelvescr
\def\scr#1{{\fam\scrfam\relax#1}}

\def\avg#1{\langle#1\rangle}
\def\Re{\mbox{Re}}
\def\Im{\mbox{Im}}
\def\Vk{\mbf k}

\begin{document}
\begin{titlepage}
\begin{flushright}
UTTG-23-97
\end{flushright}
\begin{center}
{\Large\bf A theoretical study on the damping of collective
excitations in a Bose-Einstein condensate}

\vskip 2cm
{W. Vincent Liu\footnote{liu@physics.utexas.edu} \\
\it Theory Group, Department of Physics, University of Texas, Austin, TX
78712-1081} 

\vskip 3cm
\begin{abstract}
We study the damping of low-lying collective excitations of
condensates in a weakly
interacting Bose gas model within the framework of imaginary time path
integral.  
A general expression of the damping rate has been
obtained in the low momentum limit 
for both the very low temperature regime and the
higher temperature regime. For the latter, the result is new and
applicable to  recent
experiments. Theoretical predictions for the damping rate
are compared with the experimental values.

{PACS numbers: 03.75.Fi, 67.40.Db, 11.10.Wx}
\end{abstract}
\end{center}
\end{titlepage}

The recent realization of Bose-Einstein condensation (BEC) in dilute
atomic vapors
\cite{Anderson95,Davis95} has opened the door to experimentally study 
 weakly interacting dilute quantum gases for which 
microscopic theories have been well studied for decades.   
In particular, it has been shown that the frequencies of the low-lying
collective excitations of the condensates \cite{Jin96,Mewes96} agree
excellently  with theoretical predictions based on a mean-field theory
\cite{Stringari96}. The damping in such collective modes
has been experimentally discovered by Jin {\it et al} \cite{Jin96} and
Mewes {\it et al} \cite{Mewes96}.
A very recent experiment \cite{Jin97} has extended the study of
low-lying collective excitations of condensates to include higher
temperatures in which the damping of these excitations 
exhibits dramatic temperature dependence.  So far there is no
theoretical prediction for it in such a temperature regime.

This paper aims to provide a simple way of calculating
the damping rate of collective
excitations in the above temperature regime.  
The experimental setup we are considering is a dilute gas of atoms
confined in a trap potential.
Recently,  Chou, Yang and Yu \cite{Chou96} argued that
the local density approximation can be applied into the present
problem. In other words, the trap potential may be treated as a
slowly varying external potential and enters into the theory only as a
modification to the chemical potential. To simplify the theory,
let us assume that the trap is absent in deriving the damping rate since  
we are only interesting in
finding the leading term for it.
In the following, we shall study a model of weakly interacting bosonic
particles in an infinite free space using
the imaginary time path integral method which allows us to
take into account excitation collision processes that are believed to be 
the causal of the damping.

In the low momentum  and low temperature 
limit, one usually finds
an effective field theory in which only slowly varying
degrees of freedom appear explicitly with interactions that include
the effects of fast varying fields that have been integrated out first
\cite{Weinberg96}.
One version of such an 
effective field theory has been established for a non-ideal
Bose gas in the low temperature regime by Popov \cite{Popov72,Popov83}. The
theory treats the system effectively
as that of quasiparticles described by two slowly varying real scalar
fields $\phi(x) $ and $\sigma(x)$. Here, the $\sigma(x)$ field
describes the density fluctuations, written as $\sigma(x)=n(x) -n_0$ with $n_0$
the density of the  ground state. The $\phi(x)$ field is
the Goldstone field in this theory appearing originally in the
phase of original particle fields. Notice that we have written the
four-dimensional Euclidean spacetime as $x=({\mbf x},\tau)$ with
$\tau=it$ denoting the Euclidean ``time''. According to Popov, 
the Euclidean action for such a system can be written by ($\hbar\equiv
1$ henceforth) 
\begin{eqnarray}
S[\phi,\sigma] & =& \int_0^\beta d \tau \int_{-\infty}^\infty d^3 x
\left\{ i{\partial^2 p \over \partial \mu \partial n} \sigma
\partial_\tau \phi(x)  -{1\over 2m} {\partial p\over \partial \mu}
\nabla\phi(x) \cdot\nabla\phi(x) - {1\over 2}{\partial^2 p\over \partial
\mu^2} (\partial_\tau\phi(x))^2  \right. \nonumber \\
 & & \left.
+ {1\over 2}{\partial^2 p\over
\partial n^2} \sigma^2(x) - {\nabla\sigma(x) \cdot\nabla\sigma(x)
\over 8mn_0} - {\sigma(x) (\nabla\phi(x)\cdot\nabla\phi(x)) \over 2m}
\right\}
\label{eq:Seff}
\end{eqnarray}
where $m$ is the particle mass and $p$ is the pressure of the system
given as
the function of the 
chemical potential $\mu$ and the particle density $n$.  
Notice that 
$\phi(x)$ and $\sigma(x)$  are
periodic in ``time'' $\tau$ with period $\beta=1/ (k_BT)$. 
If ${\mbf v}=(1/m)\nabla\phi$ is identified with the phonon velocity
field, one can 
easily verify that the action (\ref{eq:Seff}) corresponds to a
Hamiltonian that is one form of Landau-Khalatnikov hydrodynamic Hamiltonian
\cite{Khalatnikov89}. 
In  the low temperature regime such
that the non-condensate fraction of
particles is very small, 
the expression of the pressure $p(\mu,n)$ 
can be well approximated by that of 
$T=0$.  For a weakly interacting dilute gas, it is given by \cite{Popov-Huang}
$$p= \mu n -{t_0\over 2} n^2 $$ 
where $t_0={4\pi a \over m}$ characterizes the interaction of
particles with the $s$-wave scattering length $a$. 
It follows that 
\begin{equation}
{\partial^2 p\over \partial \mu\partial n} =1, \qquad {\partial p\over
\partial \mu } =n_0, \qquad {\partial^2 p\over \partial \mu^2} =0, \qquad
\mbox{and} \qquad {\partial^2 p\over \partial n^2} = -t_0.
\label{eq:pcoeff}
\end{equation}

\begin{figure}[htbp]
\begin{center}
\epsfig{file=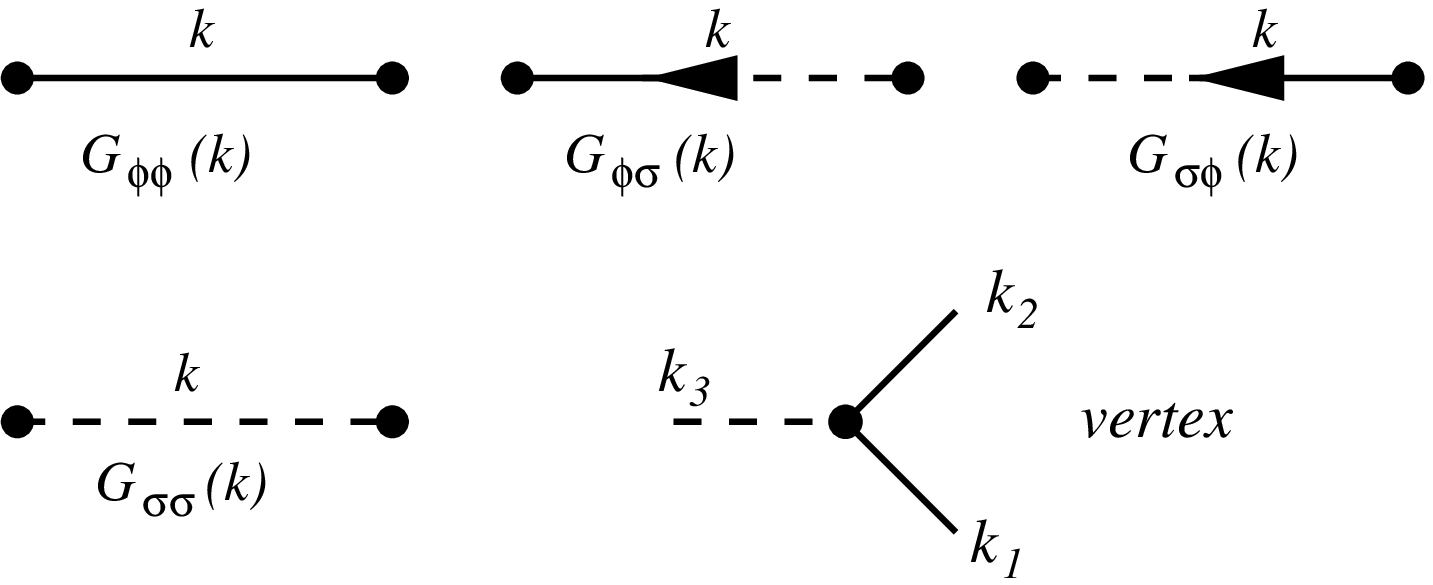,width=\linewidth}
\end{center}
\caption{Diagrams for propagators and vertex.}
\label{fig:rules}
\end{figure}

For simplicity, we write $\phi$ and $\sigma$ fields into a real
scalar doublet as $\Phi=(\phi,\sigma)$. The Green's functions (or
propagators) are defined by the matrix
$$ \displaystyle
G(x-x^\prime) =\avg{T\{\Phi(x)\Phi(x^\prime)\}} =
{\int [\prod_x d\phi(x)d\sigma(x)] \Phi(x)\Phi(x^\prime)
e^{S[\phi,\sigma]} \over \int [\prod_x d\phi(x)d\sigma(x)]
e^{S[\phi,\sigma]}} $$
with $T$ denoting a time-ordered product.
The Fourier transform of $G(x)$ is defined through
\begin{equation}
G(x) ={1 \over \beta (2\pi)^3} \sum_\nu \int d^3 k
G(k) e^{-i({\Vk} \cdot {\mbf x} -\omega_\nu
\tau)} ,
\label{eq:Fourier}
\end{equation}
where the notation $k\equiv (\Vk,i\omega_\nu)$ is understood  
and Matsubara frequencies $\omega_\nu=2\pi \nu/\beta$ $(\nu=0,\pm 1,\pm
2, \cdots)$. 
The quadratic part of the action $S$ can be written into the form of 
$S_{\mathrm{QUAD}}= \int d^4x d^4x^\prime \Phi(x) {\scr D}(x,x^\prime)
\Phi(x^\prime)$ and 
the free Green's function $G_0(k)$  is
equal to the inverse of the matrix ${\scr D}$. Thus, we find
\begin{equation}
G_0^{-1}(k) = \left(
\begin{array}{cc} \displaystyle
{n_0 \over m} k^2 & \omega_\nu \\ \displaystyle
-\omega_\nu &\displaystyle  t_0 + {k^2 \over 4mn_0}
\end{array} \right).
\label{eq:G_0inverse}
\end{equation}
It follows that
\begin{equation}
G_0(k) = \left(
\begin{array}{cc}\displaystyle
{t_0 + k^2/(4mn_0) \over \omega_\nu^2 +\epsilon^2(\Vk)} &  \displaystyle
{-\omega_\nu \over \omega_\nu^2 +\epsilon^2(\Vk)} \\ \displaystyle
{\omega_\nu \over \omega_\nu^2 +\epsilon^2(\Vk)} &\displaystyle
{(n_0/m) k^2 \over \omega_\nu^2 +\epsilon^2(\Vk)}
\end{array} \right),
\label{eq:G_0}
\end{equation}
where the spectrum
\begin{equation}
\epsilon({\Vk}) = \sqrt{({k^2\over 2m})^2 + c^2k^2}
\end{equation}
with $c\equiv \sqrt{t_0n_0/m}$. 
The propagators $G_0$ are represented by Feynman diagrams in
Fig.~\ref{fig:rules}.
Further, the cubic term of the action
(\ref{eq:Seff}) describes the interaction of three excitations that
are known as phonons in the low momentum region, giving rise to  
a vertex of $$\delta^3({\Vk_1 +\Vk_2 +\Vk_3})  
\delta_{\nu_1+\nu_2+\nu_3, 0} {({\Vk_1 \cdot \Vk_2}) \over m}$$
represented by the last diagram of Fig.~\ref{fig:rules}.
Obviously, it can be treated as perturbation in the low momentum limit.

\begin{figure}[htbp]
\begin{center}
\epsfig{file=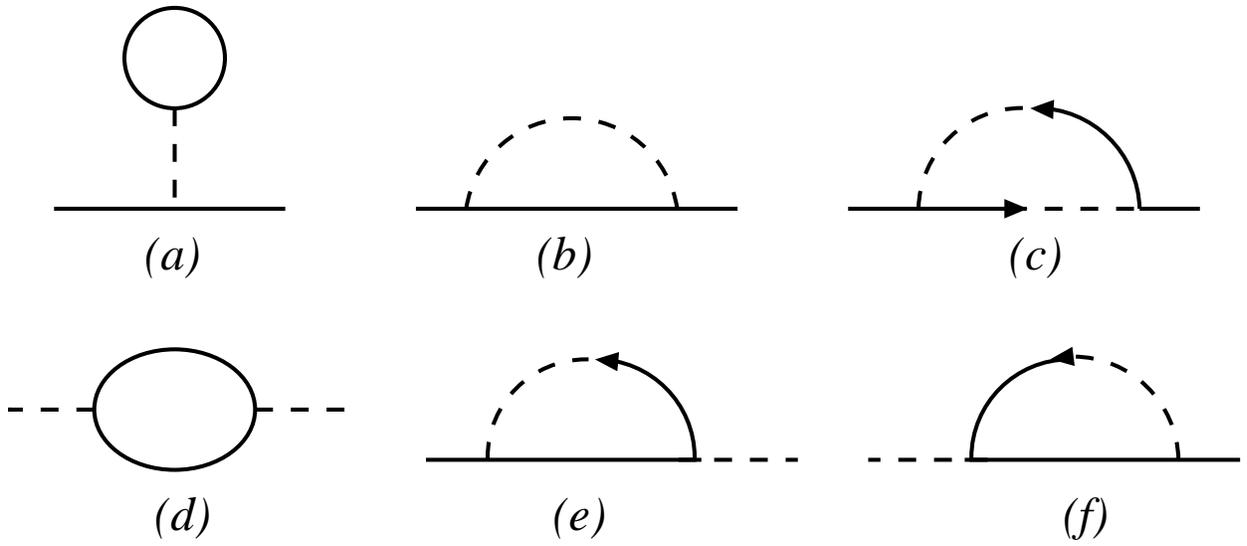,width=\linewidth}
\end{center}
\caption{One-loop diagrams for the self-energy matrix $\Pi(k)$.} 
\label{fig:loop}
\end{figure}

The spectrum of collective excitations is given by the poles of the
exact Green's function \cite{Fetter71}
$G(k)=G_0(k) + G_0(k)\Pi(k) G(k)$
where $\Pi(k)$ denotes the matrix of the self-energy parts.
It follows that the spectrum is determined by
\begin{equation}
\det G^{-1}(k) = \det(G_0^{-1}(k) -\Pi(k)) =0.
\label{eq:det}
\end{equation}
Now that we are working in the imaginary time formalism, 
let us  make the analytical continuation $i \omega_\nu \rightarrow \omega
+i\eta$ ($\eta\equiv 0^+$) after the Matsubara frequency sum 
and write $\omega =\Re\omega -i
\gamma(\Vk)$ with 
$\gamma$ denoting the damping rate. Then, keeping only terms up to one-loop
order \cite{remark:gamma}, we find from
Eqs.~(\ref{eq:G_0inverse}) and (\ref{eq:det}) that 
\begin{equation}
\gamma(\Vk)= {1\over 2\Re\omega} 
\left[(t_0+{k^2\over 4mn_0})\Im\Pi_{\phi\phi}(\Vk,\omega +i\eta) + 
{n_0k^2 \over m}\Im\Pi_{\sigma\sigma}(\Vk,\omega+i\eta)
\right ] - \Re\Pi_{\phi\sigma}(\Vk,\omega+i\eta)
\label{eq:gamma}
\end{equation}
with the understanding that all $\omega$'s are
replaced with $\epsilon(\Vk)$ after the
analytical continuation.
Up to one-loop order, there are six diagrams (see Fig.~\ref{fig:loop})
for the self-energy matrix $\Pi(\Vk, \omega)$
but only the last five (b)-(f) contribute to the damping.
After  collecting
contributions from all related diagrams, we have
\begin{equation}
\gamma(\Vk) =\gamma_1(\Vk) + \gamma_2(\Vk),
\label{eq:gammaA}
\end{equation}
with
\begin{eqnarray}
\gamma_1(\Vk) &=& {1\over 32\pi^2}\int d^3 k^\prime
\delta(\epsilon(\Vk)-\epsilon(\Vk^\prime) -\epsilon(\Vk-\Vk^\prime))
[f(\epsilon(\Vk^\prime)) -f(-\epsilon(\Vk-\Vk^\prime))]\times
\nonumber\\
 && \left\{
{(\Vk-\Vk^\prime)^2 (\Vk\cdot\Vk^\prime)^2 \epsilon(\Vk^\prime) \epsilon(\Vk)
\over 2mn_0 k^2{k^\prime}^2\epsilon(\Vk-\Vk^\prime)}  +
{(\Vk\cdot\Vk^\prime) (\Vk\cdot(\Vk-\Vk^\prime))\epsilon(\Vk)  
\over 2mn_0 k^2}
 \right. \nonumber \\
 &&  \left. 
+{k^2 (\Vk^\prime\cdot(\Vk-\Vk^\prime))^2 
\epsilon(\Vk^\prime)\epsilon(\Vk-\Vk^\prime) \over 4 mn_0 
{k^\prime}^2 (\Vk-\Vk^\prime)^2\epsilon(\Vk)}
+
{(\Vk^\prime\cdot(\Vk-\Vk^\prime))(\Vk\cdot\Vk^\prime)\epsilon(\Vk^\prime)
\over mn_0{k^\prime}^2} \right\}
\end{eqnarray}
and
\begin{eqnarray}
\gamma_2(\Vk) &=& {1\over 32\pi^2}\int d^3 k^\prime
\delta(\epsilon(\Vk)+\epsilon(\Vk^\prime) -\epsilon(\Vk+\Vk^\prime))
[f(\epsilon(\Vk^\prime)) -f(\epsilon(\Vk+\Vk^\prime))]\times
\nonumber\\
 && \left\{ {\epsilon(\Vk) \over 2mn_0} \left[  
{{k^\prime}^2 (\Vk\cdot(\Vk+\Vk^\prime))^2 \epsilon(\Vk+\Vk^\prime) 
\over  k^2(\Vk+\Vk^\prime)^2\epsilon(\Vk^\prime)}  + 
{(\Vk+\Vk^\prime)^2 (\Vk\cdot\Vk^\prime)^2 \epsilon(\Vk^\prime) 
\over  k^2{k^\prime}^2\epsilon(\Vk+\Vk^\prime)} \right] \right. \nonumber \\
 && +{(\Vk\cdot\Vk^\prime) (\Vk\cdot(\Vk+\Vk^\prime))\epsilon(\Vk)  
\over mn_0 k^2} 
+{k^2 (\Vk^\prime\cdot(\Vk+\Vk^\prime))^2 
\epsilon(\Vk^\prime)\epsilon(\Vk+\Vk^\prime) \over 2 mn_0 
{k^\prime}^2 (\Vk+\Vk^\prime)^2\epsilon(\Vk)} \nonumber \\
&&  \left. 
   + {(\Vk^\prime\cdot(\Vk+\Vk^\prime))\over mn_0} \left[
  {(\Vk\cdot\Vk^\prime)\epsilon(\Vk^\prime) \over {k^\prime}^2} + 
  {(\Vk\cdot(\Vk+\Vk^\prime))\epsilon(\Vk+\Vk^\prime) \over
(\Vk+\Vk^\prime)^2 } \right] \right\},
\end{eqnarray}
where $f(\epsilon)=1/(\exp(\beta \epsilon) -1)$.

If $T=0$, one can easily verify that for small $k$ ($ck \ll n_0t_0$)
\begin{equation}
\gamma_{T=0}(\Vk) = {3k^5\over 640\pi mn_0}  \label{eq:gammaT=0}
\end{equation}
which is the well-known Beliaev's result \cite{Beliaev58}.
The product $n_0t_0$  characterizes  the strength of
particle interactions. 

For $T\neq 0$ and small $k$ such that  $ck \ll k_BT$ and $ck \ll
n_0t_0$, we find  that
the damping rate to the lowest order in $k$ is determined by
\begin{equation}
\gamma(\Vk) = {ckk_0^5 \over 16\pi mn_0 k_BT} {\scr I}
({n_0t_0\over k_BT}) \qquad 
\label{eq:gammaB}
\end{equation}
where $k_0\equiv \sqrt{mn_0t_0}$ and  the function ${\scr I}(x)$ is given by
$$
{\scr I}(x)={\pi^2\over 6 x^3} + \int_0^\infty d\xi \left[ {2\xi^2 \over
(1+\xi^2)^{3/2} } + {3\over 2(1+\xi^2)} - {2\over (1+\xi^2)^2} \right] 
{\xi^2 e^{x \xi} \over (e^{x \xi}-1)^2} \quad .
$$
If the limit ${k_BT \over n_0t_0}\rightarrow 0$ is taken, Eq.~(\ref{eq:gammaB})
reduces to  the familiar form
\begin{equation}
\gamma(\Vk) = {3\pi^3 k(k_BT)^4 \over 40mn_0c^4} \label{eq:gammaTlow}
\end{equation}
which was given by Hohenberg and Martin \cite{Hohenberg65}.
Though, we shall see in the following paragraph that
Eq.~(\ref{eq:gammaTlow}) is invalid in the recent experimental
temperature regime. 
In fact, both expressions (\ref{eq:gammaT=0}) and
(\ref{eq:gammaTlow}) were already derived by Popov in
Ref.~\cite{Popov72} from the effective action (\ref{eq:Seff}). 
But, to the best of my knowledge, the general expression
(\ref{eq:gammaB}) is first obtained here.
It is valid as long as the system is in the low temperature region
such that the number of
particles in the excited states is much less than that in the condensate.

\begin{figure}[htbp]
\begin{center}
\epsfig{file=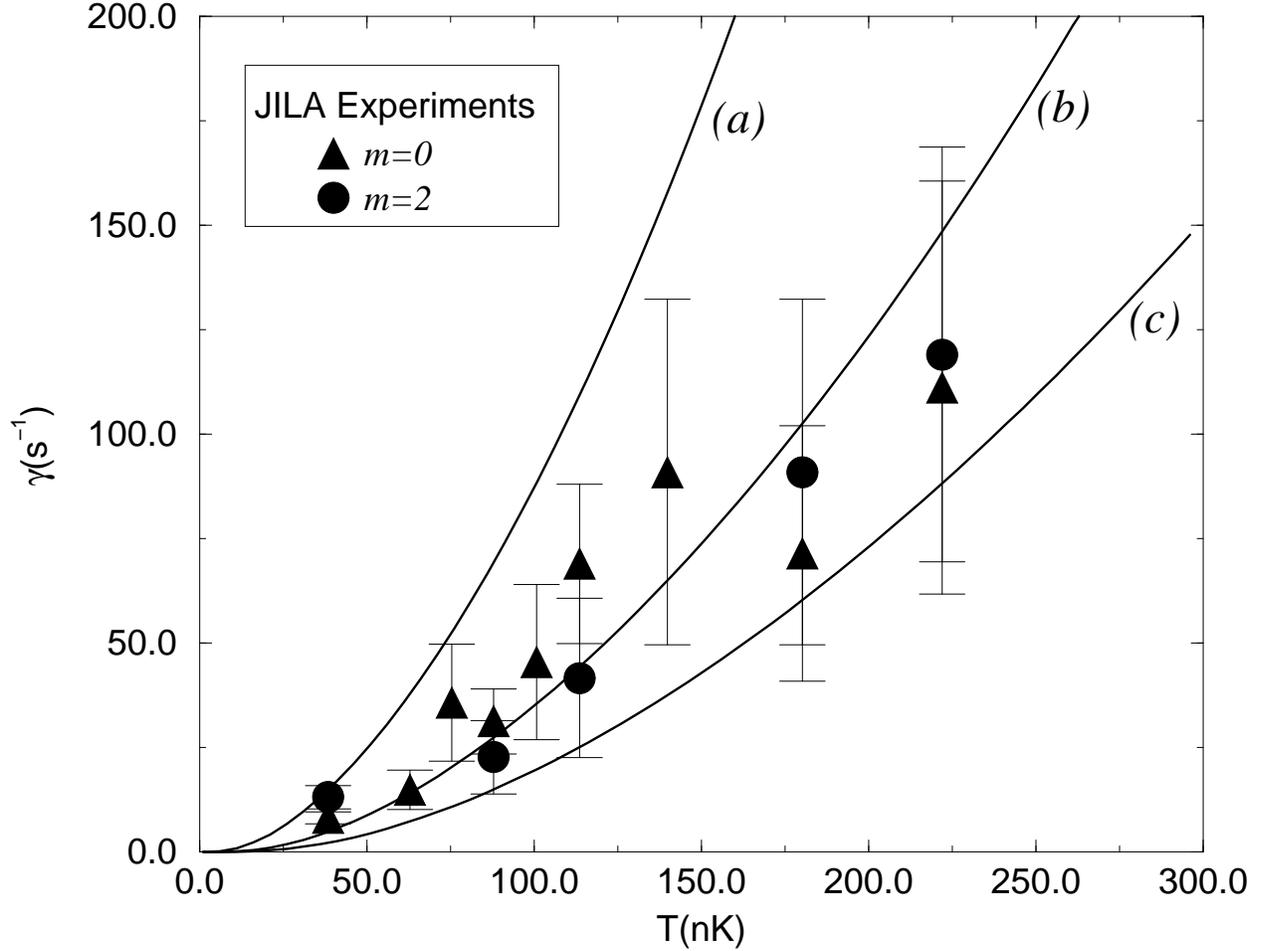,width=\linewidth}
\end{center}
\caption{The damping rate of collective excitations in a $^{87}$Rb
atomic gas. Three solid lines indicate theoretical predictions from
Eq.~(\ref{eq:gammaB}). For those, the excitation frequency is taken as
$\omega/2\pi=1.4\nu_r=180.6$Hz corresponding to the mode $m=2$ and the
condensate densities are: $(a)$ $n_0=1.0\cdot 10^{14}$cm$^{-3}$; $(b)$
$n_0=2.0\cdot 10^{14}$cm$^{-3}$; and $(c)$ $n_0=3.0\cdot
10^{14}$cm$^{-3}$. Two discrete curves are replotted from the data of
Ref.~\protect\cite{Jin97}.  (The experimental data is the courtesy of
M. R. Matthews \protect\cite{Matthews97}.) }

\label{fig:dp_jila}
\end{figure}

The damping of collective excitations in BEC has been measured in the
dilute atomic vapors of both $^{87}$Rb  and sodium at JILA  and MIT,
respectively \cite{Jin96,Mewes96,Jin97}.
For the experiment of $^{87}$Rb at JILA, the condensates are produced
in a trap with frequencies of $\nu_r=129$Hz radially and $\nu_z=365$Hz
axially. For a typical condensate of $N_{\mathrm{BEC}}=4500$ atoms, the
density of a 
condensate can be estimated as $n_0 \approx 1.4\times
10^{14}$cm$^{-3}$ \cite{Baym96}. Further, the scattering length
for a $^{87}$Rb vapor may be taken roughly as $a=103$Bohr \cite{Cornell97}.
It turns out that the typical interaction energy $n_0t_0 \approx
57$nK. Therefore, 
the damping rate is expected to be given by Eq.~(\ref{eq:gammaB})
instead of Eq.~(\ref{eq:gammaTlow}) in the temperature region ($30,
300$)nK where it was measured (see Figs.~1 and 3 of
Ref.~\cite{Jin97}). 
We also checked that for the lowest excitation mode ($m=2$) both
${ck/ (k_BT)}$ 
and ${ck/( n_0t_0)}$ are less than $0.2$ for any $T$ above $50$nk.
Fig.~\ref{fig:dp_jila} shows  theoretical predictions for the damping
rate of collective excitations based on Eq.~(\ref{eq:gammaB}) in
comparison with
the experimental data from Ref.~\cite{Jin97}. 
The theoretical curve $(b)$ seems to fit the
experiment well 
apart from that the theoretical values are varying with the density of
condensates approximately according to $\gamma \propto 1/\sqrt{n_0^3}$. 
However, a Bose condensate in a trap is essentially inhomogeneous
and we will not try to determine a proper average density $n_0$ here. 
Next, let us check whether the above experiment is in the low
temperature region. 
For $T<100$nk, we read off from Fig.~1 of Ref.~\cite{Jin97}  that the
corresponding $T^\prime <0.6$ (where $T^\prime\equiv T/T_c$) 
and the condensate fraction $N_{\mathrm{BEC}}/N$ is greater than or
about $80\%$. 
Hence, the higher order corrections to Eq.~(\ref{eq:gammaB}) due to 
finite temperature effects only can be
estimated from the calculation of Feynman diagrams less than
$(1-N_{\rm BEC}/N)^2=4\%$. 
On the other hand side,  Ref.~\cite{Jin97} shows that more
than $50\%$ of particles reside in noncondensates when $T^\prime>0.8$ and 
correspondingly $T>175$nK roughly. This means that the damping rate
shall be no longer determined by  Eq.~(\ref{eq:gammaB}) in that
temperature regime. I do not know yet whether there exists any
theoretical prediction for it.

\begin{figure}[htbp]
\begin{center}
\epsfig{file=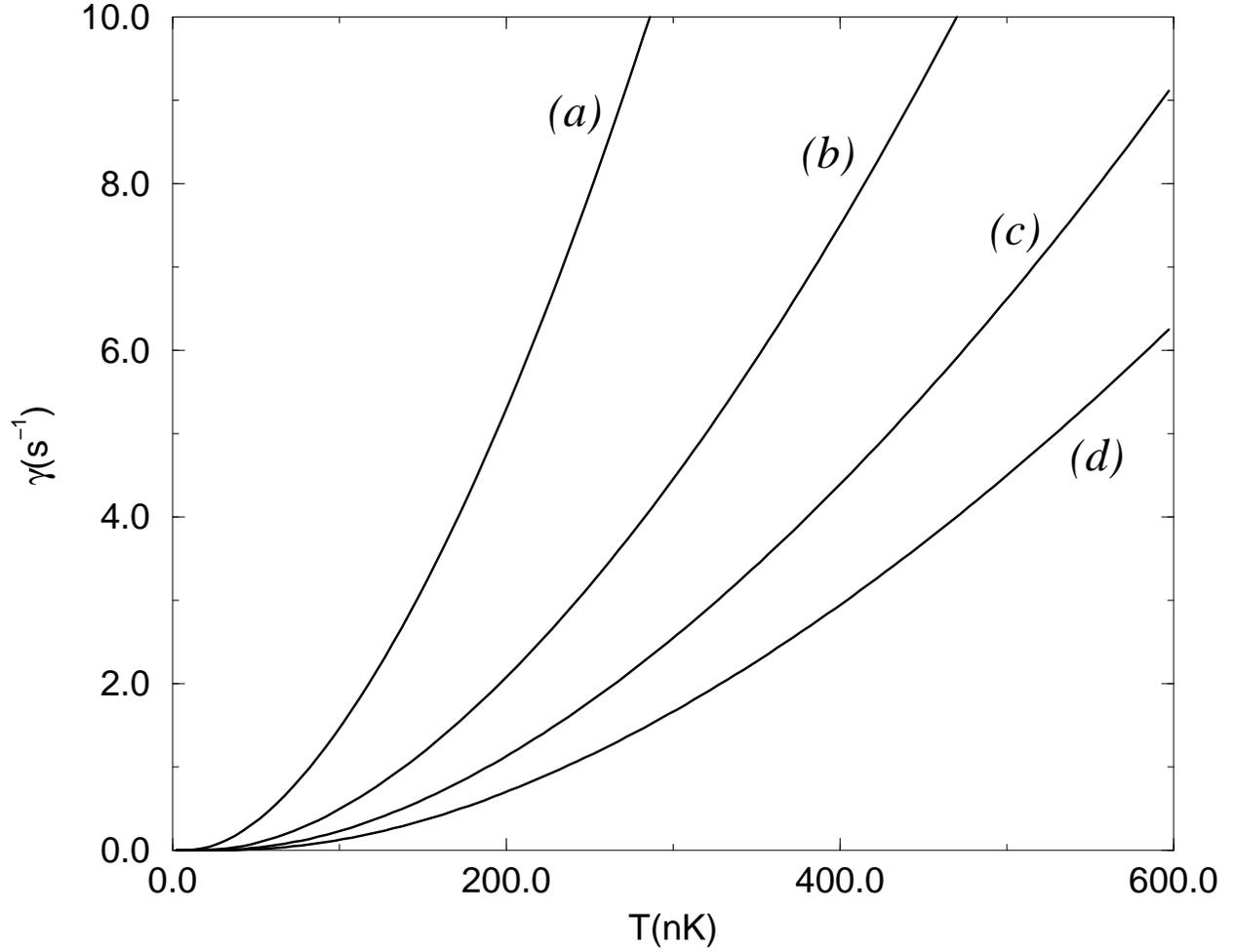,width=\linewidth}
\end{center}
\caption{The damping rate of collective excitations in a sodium atomic
gas. All solid
lines are plotted according to Eq.~(\ref{eq:gammaB}).
The excitation frequency is $30$Hz and the condensate densities are:
$(a)$  $n_0=1.0\cdot 10^{14}$cm$^{-3}$; 
$(b)$  $n_0=2.0\cdot 10^{14}$cm$^{-3}$; 
$(c)$  $n_0=3.0\cdot 10^{14}$cm$^{-3}$; and 
$(d)$  $n_0=4.0\cdot 10^{14}$cm$^{-3}$. }
\label{fig:dp_mit}
\end{figure}

In Fig.~\ref{fig:dp_mit}, we plot the damping rate of the collective
excitation of frequency $30$Hz for  the sodium gas system of
Mewes {\it et al} \cite{Mewes96} using  
the $s$-wave scattering length $a=65$Bohr \cite{Mewes9607}. 
The decay time of $250(40)$ms was found experimentally when a nearly
pure condensate was formed at the
temperature $T\approx 0.5T_c$ \cite{Ketterle97}.
The Bose-Einstein transition temperature is determined theoretically according
to $T_c= {\hbar\bar{\omega}\over k_B}(N/1.202)^{1/3}$
\cite{deGroot50,Mewes9607} with $\bar{\omega}$ the geometric mean of
the harmonic trap frequencies
$\bar{\omega}=(\omega_x\omega_y\omega_z)^{1/3}$. For that the trap used
in Ref.~\cite{Mewes96} is of frequencies of $250$Hz radially and $19$Hz
axially and typically holds a total number $N \approx 5\times 10^{6}$
of atoms when the decay time was measured, we have $T_c\simeq 800$nK.
For a typical condensate density $n_0=3\times 10^{14}$cm$^{-3}$
\cite{Mewes9607}, we can read off from the curve $(c)$ of
Fig.~\ref{fig:dp_mit} that $\gamma\simeq 4.4$s$^{-1}$ for  temperature
$T=400$nK. This damping rate corresponds to a decay time of about
$230$ms that agrees very well with the experimental value.  Also,
since at such a temperature the condensate fraction of atoms is around
$90\%$, higher order corrections to $\gamma$ due to the finite
temperature effect alone are quite small ($<1\%$). Also, we checked
that both ${ck/( k_BT)}$ 
and ${ck/ (n_0t_0)}$ are less than one percent. That is to say, the
system is well in the low momentum region.

In conclusion, 
this paper analytically calculates the damping rate of collective
excitations for a dilute Bose gas model in a temperature regime where
theoretical predictions did not exist previously.  
Although the model has ignored the contribution of the trap and has
required that the condensate be homogeneous, it produces  results for
the damping of collective excitations 
that are in good agreement with the experiments.
Our study also reveals that the damping is due to the
three-phonon collision processes at finite temperature. 
This paper serves a very preliminary study of this subject and fully
understanding it requires a complete analysis to take into
account the contribution of the trap and the inhomogeneity of the
system.

I am indebted to Professor S. Weinberg for guidances and
carefully reading the manuscript. 
I am  grateful to E. Cornell and  W. Ketterle
for  very helpful instructions on the experiments, to
M. R. Matthews for sending me the experimental data, to G. Ordonez for
helpful conversations, and to Professor E. C. G. Sudarshan for
illuminating discussions on BEC.
This work is supported by NSF grant PHY-9511632 and the
Robert A. Welch Foundation.

\newpage

\begin{thebibliography}{99}
\bibitem{remark:gamma} We
assume that proper renormalization
procedures have been carried out so that no diagram for the matrix
$\Pi$ diverges.
\bibitem{Ketterle97} W. Ketterle pointed out to me that a ``nearly
pure condensate'' meant that the condensate fraction of atoms was
greater than or about $90\%$. Based on this condition and 
 $N_{\mathrm{BEC}}/N=1-(T/T_c)^3$ \cite{deGroot50}, he argued that the absolute
temperature was about $0.5T_c$.
\end{thebibliography}
\end{document}